# Investigation of the Ground and Excited States of the $^3$H and $^4$He Nuclei by Using the Methods of Group Theory

**S. B. Doma[1) and H. S. El-Gendy[2)**

[1)] Faculty of Science, Alexandria University, Alexandria, Egypt, sbdoma@yahoo.com
[2)] Faculty of Science & Art, Shaqra University, Shaqra, KSA, hselgendy1@yahoo.com

**Abstract**
The method of group theory is applied to investigate the ground- and the excited states of the $^3$H and $^4$He nuclei by using the translation invariant shell model with basis functions corresponding to even number of quanta of excitations in the range $0 \leq N \leq 20$. Accordingly, the ground and first-excited state wave functions and energies, the S-, P- and D-state probabilities, the root mean-square radius and the magnetic dipole moment of triton have been investigated. Also, the energies and wave functions of the ground and the even-parity excited states and the root mean-square radius of $^4$He have been investigated. Two residual two-body interactions together with two three-nucleon interactions have been used in the calculations. Moreover, the convergence of calculations has been examined by extrapolating the results with $N \leq 20$ step-by-step to reach $N = 30$ for the two nuclei.

**1. Introduction**
The nuclear shell model [1] has achieved wide popularity among nuclear theorists owing to its easily visualizable character and the success in interpreting experimental results. The assumption that, to a first approximation, each nucleon moves in an average potential independent of the motion of the other nucleons is an attractive one, due to one's familiarity with the Hartree fock theory of atomic structure.

The methods of expanding the nuclear wave function in terms of a complete set of orthonormal functions, basis functions, have been used on a large scale especially for the nuclei with $3 \leq A \leq 6$ [2].

The translation invariant shell model [2-5] (TISM), also known as the unitary scheme model (USM) due to the main role of the unitary group in the construction of the nuclear wave functions, has shown good results for the structure of light nuclei with $A \leq 7$ by using nucleon-nucleon interactions [5-10]. This model considers the nucleus as a system of noninteracting quasi particles and enables us to apply the algebraic methods for studying the general features of matrix elements of operators that correspond to physical quantities. The TISM is based on the group theoretical methods of classifying the basis functions. The basis functions of this model are constructed in such a way that they will have certain symmetry with respect to the interchange of particles and have definite total angular momentum $J$ and isotopic spin $T$. The basis functions of the TISM are then expanded in the form of products of two types of functions, one corresponds to the set of A-2 nucleons and the other corresponds to the last pair of nucleons by means of the two-particle fractional parentage coefficients [2,6,9]. Using two-body interactions it is then possible to calculate the Hamiltonian matrices of the different nuclear states. In principle, the predicted results for the nuclear characteristics should be independent of the particular chosen basis functions when the number of terms in the expansion is kept large enough. The inclusion of all such bases in the expansion is too difficult since the matrices of the two-particle fractional parentage coefficients corresponding to these basis functions grow rapidly. It is therefore fundamental to have some rules that would allow us to reduce the number of these bases. Some of these rules are adopted for light nuclei in ref. [2] and in ref. [6] for the nuleus $^6$Li.

On the other hand, recent experimental results in three-body systems have unambiguously shown that calculations based only on nucleon–nucleon forces fail to accurately describe many experimental observables and one needs to include effects which are beyond the realm of the two-body potentials. Also, microscopic calculations of light nuclei and nuclear matter [11] have indicated that it is difficult to explain the observed binding-energies and densities if we assume a non relativistic nuclear Hamiltonian having only two-nucleon interactions, consistent with the nucleon-nucleon scattering data at low energies ($E_{lab} < \sim 400 MeV$).

Since nucleons are composite objects made up of quarks and gluons, we can not approximate their interactions by a sum of two-body terms. The mesonic degrees of freedom can also generate three- and more-body potentials in the Hamiltonian in which only the nucleon degrees of freedom are retained. Since the energies obtained with Hamiltonian having only two-body potentials are not far from the experiment, we expect that the contribution of the many-body potentials is small compared to that of the two-body interaction in the realm of nuclear physics, and particularly, only three body-potentials may be important.

The effect of the three-nucleon interactions has been studied recently [12], where the effect of the different three-nucleon interactions in $p$-$^3$He elastic scattering at low energies has been calculated for the four-nucleon scattering observables by using the Kohn variational principle and the hyperspherical harmonics technique. On the other hand, the effects of the two-body and the three-body hyperon-nucleon interactions in $\Lambda$ hypernuclei have been stydied by assessing the relative importance of two- and three-body hyperon-nucleon force and by studying the effect of the hyperon-nucleon-nucleon interaction in closed shell $\Lambda$ hypernuclei from A = 5 to 91 [13,14]. Moreover, Cipollone et al. [15] extended the formalism of self-consistent Green's function theory to include three-body interactions and applied it to isotopic chains around oxygen for the first time. Furthermore, Wiringa et al. [16] used the realistic Argonne $v_{18}$ potential for the two-nucleon interaction and Urbana three-nucleon potentials to generate accurate variational Monte Carlo (VMC) wave functions for the $A \leq 12$ nuclei.

The ab initio no-core shell model (NCSM) is a well-established theoretical framework aimed at an exact description of nuclear structure starting from high-precision interactions between the nucleons. Barrett, Navrátil, and Vary [17] discussed, in details, the extension of the ab initio NCSM to nuclear reactions and sketched a number of promising future directions for research emerging from the NCSM foundation, including a microscopic non-perturbative framework for the theory with a core. In the NCSM, Forssén, Navrátil and Quaglioni [18] considered a system of A point-like non-relativistic nucleons that interact by realistic inter-nucleon interactions. They considered two-nucleon interactions that reproduce nucleon-nucleon phase shifts with high precision, typically up to 350 MeV lap energy. Also, they included three-nucleon interactions with terms, e.g., related to two-pion exchanges with an intermediate delta excitation. Both semi-phenomenological potentials, based on meson-exchange models, as well as modern chiral interactions are considered.

It is well-known that calculations within a three-body translational invariant harmonic oscillator basis and using realistic two and three-nucleon forces have been performed for the three-nucleon system (see e.g. [19]), and for the four-nucleon system (see e.g. [20,21]). These were obtained within the translational invariant form of the no-core shell model, which is equivalent to the TISM with the exception that the antisymmetrization of the wave function is not achieved by means of group theory, but rather by diagonalizing the antisymmetrization operator and retaining as basis states the antisymmetric eigenstates.

It is well-known also that, the three- and four-nucleon systems have been studied by means of numerically exact few-body approaches (such as the Faddeev, Faddeev-Yacubosky and hyperspherical harmonics approaches starting from realistic two and three-nucleon forces, see for example [22-24]). A review can be found e.g., in [25].

In previous papers Doma et al. [9] applied the TISM with number of quanta of excitations $0 \leq N \leq 8$ to investigate the ground-state wave function, the binding energy, the first excited state energy and the root mean-square radius of $^3$H by using the Gogny, Pires and De Tourreil (GPT) interaction [26], the Hu and Massey potential [27] and an effective interaction proposed by Vanagas [2]. Furthermore, Doma [10] applied the TISM with number of quanta of excitations $0 \leq N \leq 8$ to investigate the ground- and excited-states wave functions, the spectrum, the root mean-square radius and the integral cross section of the $\gamma$-quanta photo absorption of the nucleus $^4$He by using the GPT-potential.

In the present paper we have applied the TISM with basis functions corresponding to even number of quanta of excitations in the range $0 \leq N \leq 20$ to investigate the ground and first-excited state wave functions and energies, the S-, P- and D-state probabilities ($P_S, P_P, P_D$), the root mean-square radius and the magnetic dipole moment of triton. Also, the wave functions and energies of the ground and the even-parity excited states and the root mean-square radius of $^4$He have been investigated. In carrying out the calculations in these investigations we have used two nucleon-nucleon interactions and two three-nucleon interactions. The two nucleon-nucleon interactions (Pot-I and Pot-II) have the same shape and structure as that proposed by Doma et al. [28]; i.e. they consist of central, tensor, spin-orbit and quadratic spin-orbit forces with Gaussian radial dependences, which are extremely suitable for the calculations of the different matrix elements with respect to the harmonic-oscillator basis functions. New parameters for these potentials are calculated for the present paper and are so chosen in such a way that they represent the long-range attraction and the short-range repulsion of the nucleon-nucleon interactions. These parameters are also chosen so as to produce good agreement between the calculated values of the binding energy, the root mean-square radius, the D-state probability, the magnetic dipole moment and the electric quadrupole moment of deuteron, as for the original potential [28]. For the three-body interactions we have used two potentials; the first one is of the form of Skyrme III potential [29] and the second one is the Urbana model (UVII) three nucleon interaction, which is given in the form of a sum of long-range two pion exchange and intermediate-range repulsive terms [30]. Moreover, the convergence of the calculations with the TISM has been examined by extrapolating the results with $N \leq 20$ step-by-step to reach $N = 30$ for the two nuclei.

## 2. The Translation Invariant Shell Model

The Hamiltonian $\mathcal{H}$ of a nucleus consisting of A-nucleons, interacting via two-body potential, can be written in terms of the relative coordinates of the nucleons, in the form [10]

$$\mathcal{H} = \frac{1}{2m}\sum_{i=1}^{A} \boldsymbol{p}_i^2 + \frac{1}{2}\sum_{i=1\neq j}^{A}\sum_{j=1}^{A} V(|\boldsymbol{r}_i - \boldsymbol{r}_j|). \qquad (2.1)$$

The translational invariance of the Hamiltonian $\mathcal{H}$ permits the separation of the center of mass motion, and as a consequence the Hamiltonian corresponding to the internal motion becomes

$$H = \mathcal{H} - T_{CM}, \tag{2.2}$$

where $T_{CM}$ denotes the center of mass kinetic energy

$$T_{CM} = \frac{1}{2mA}\mathbf{P}^2 = \frac{1}{2mA}\left(\sum_{i=1}^{A}\mathbf{p}_i\right)^2. \tag{2.3}$$

By adding and subtracting an oscillator potential referred to the center of mass, the internal Hamiltonian $H$ becomes

$$H = \frac{1}{2m}\sum_{i=1}^{A}\mathbf{p}_i^2 - \frac{1}{2mA}\mathbf{P}^2 + \frac{1}{2}m\omega^2\sum_{i=1}^{A}(\mathbf{r}_i - \mathbf{R})^2 + \frac{1}{2}\sum_{i=1\neq j}^{A}\sum_{j=1}^{A}V(|\mathbf{r}_i - \mathbf{r}_j|)$$

$$-\frac{1}{2}m\omega^2\sum_{i=1}^{A}(\mathbf{r}_i - \mathbf{R})^2. \tag{2.4}$$

The internal Hamiltonian can be rewritten in terms of the relative coordinates of the nucleons in the form

$$H = H^{(0)} + V' \tag{2.5}$$

where

$$H^{(0)} = \frac{1}{2m}\sum_{i=1}^{A}\mathbf{p}_i^2 - \frac{1}{2mA}\mathbf{P}^2 + \frac{1}{2}m\omega^2\sum_{i=1}^{A}(\mathbf{r}_i - \mathbf{R})^2$$
$$= \frac{1}{A}\sum_{1=i<j}^{A}\left[\frac{1}{2m}(\mathbf{p}_i - \mathbf{p}_j)^2 + \frac{1}{2}m\omega^2(\mathbf{r}_i - \mathbf{r}_j)^2\right], \tag{2.6}$$

is the well-known TISM Hamiltonian, also known as the USM Hamiltonian, and

$$V' = \sum_{1=i<j}^{A}\left[V(|\mathbf{r}_i - \mathbf{r}_j|) - \frac{m\omega^2}{2A}(\mathbf{r}_i - \mathbf{r}_j)^2\right]. \tag{2.7}$$

is the residual two-body interaction.

The energy eigenfunctions and eigenvalues of the Hamiltonian $H^{(0)}$ are given by [10]

$$|A\,\Gamma; M_L M_S T M_T\rangle \equiv |A\,N\{\rho\}(\nu)\alpha[f]LS; M_L M_S T M_T\rangle, \tag{2.8}$$

$$E_N^{(0)} = \left\{N + \frac{3}{2}(A-1)\right\}\hbar\omega. \tag{2.9}$$

In eq. (2.8) $\Gamma$ is the set of orbital and spin-isospin quantum numbers $N\{\rho\}(\nu)\alpha[f]LS$ characterizing the states. The number of quanta of excitations $N$ is an irreducible representation (IR) of the unitary group $U_{3(A-1)}$. The representation $\{\rho\} = \{\rho_1, \rho_2, \rho_3\}$, $\rho_1 + \rho_2 + \rho_3 = N$, is related to Elliott symbol $(\lambda\mu)$ [31] by the relations: $\lambda = \rho_1 - \rho_2$, $\mu = \rho_2 - \rho_3$. $\{\rho\}$ is an IR of the unitary group $U_{(A-1)}$ and the unitary unimodular subgroup of three dimensions $SU_3$, at the same time. The representation $(\nu)$ is an IR of the orthogonal group $O_{A-1}$ and $[f]$ is an IR of the symmetric group $S_A$. $L$ and $M_L$ stand for the orbital angular momentum and its z-component and are also IR of the rotational groups $SO_3$ and $SO_2$, respectively. $S, M_S$ are the spin, its z-component and $T, M_T$ are the isospin, its z-component which are IR of the direct product of the groups $SU_2 \dot\times SU_2$. $\alpha$ is a repetition quantum number. The functions (2.8) form a complete set of functions, bases. It is easy to construct bases which have definite total momentum $J$ in the form [8,10]

$$|A \Gamma JM_J TM_T\rangle = \sum_{M_L+M_S=M_J}(LM_L,SM_S|JM_J)|A \Gamma; M_L M_S TM_T\rangle, \qquad (2.10)$$

where $(LM_L,SM_S|JM_J)$ are the Clebsch-Gordan coefficients of the rotational group $SO_3$. The nuclear wave function of a state with total momentum $J$, isospin $T$ and parity $\pi$ can be constructed as follows [8,10]

$$|A J^\pi TM_J M_T\rangle = \sum_\Gamma C_\Gamma^{J^\pi T}|A \Gamma JM_J TM_T\rangle, \qquad (2.11)$$

where $C_\Gamma^{J^\pi T}$ are the state-expansion coefficients. In the summation (2.11) the number of quanta of excitations $N$ is permitted to be either even or odd integer depending on the parity of the state $\pi$.

The matrix elements of the residual two-body interaction $V'$, Eq. (2.7), with respect to the bases (2.8) are given in details in [8,10,32,33] by expanding these bases in the form of products of two types of functions: the first of which corresponds to the set of $A-2$ nucleons and the second corresponds to the set of the last two nucleons in terms of the two-particle fractional parentage coeeficients, which are products of orbital and spin-isospin two-particle fractional parentage coefficients. The ground-state nuclear wave function, which is obtained as a consequence of the diagonalization of the ground-state energy matrix, is used to calculate the root mean-square radius and the magnetic dipole moment.

## 3. The Root Mean-Square Radius and the Magnetic Dipole Moment
The root mean-square radius is defined by

$$\mathcal{R} = \sqrt{r_p^2 + \langle R_{Nuc}^2\rangle}, \qquad (3.1)$$

where $r_p = 0.85$ fm is the proton radius, and the second term is the mean value of the operator [6]

$$R_{Nuc}^2 = \frac{1}{A^2}\sum_{1=i<j}^A r_{ij}^2. \qquad (3.2)$$

This operator does not depend on the spin-isospin variables of the nuclear wave function and its calculation is straightforward [6].

The nuclear magnetic dipole moment is defined as the mean value of the operator $\hat{\mu} = \hat{\mu}_\sigma + \hat{\mu}_0$, where [2]

$$\hat{\mu}_\sigma = \sum_{i=1}^A[(\mu_p+\mu_n) + 2(\mu_p-\mu_n)t_{0i}]s_{0i}, \qquad (3.3)$$

and

$$\hat{\mu}_0 = \frac{1}{2}\sum_{i=1}^A[(1-2t_{0i})]\ell_{0i}, \qquad (3.4)$$

calculated in a state with $M_J = J$. In euations (3.3) and (3.4) $\mu_p$ and $\mu_n$ are the proton and neutron magnetic moments, respectively. $t_{0i}$, $s_{0i}$ and $\ell_{0i}$ are the z-components of the isospin, spin and orbital momenta of the $i$th nucleon, respectively. Writing each of the two operators $\hat{\mu}_\sigma$ and $\hat{\mu}_0$ as a sum of symmetric and antisymmetric operators of symmetry-types [A] and [A-1,1], in the form

$$\hat{\mu}_\sigma = \mu_\sigma^{[A]} + \mu_\sigma^{[A-1,1]}, \quad \hat{\mu}_0 = \mu_0^{[A]} + \mu_0^{[A-1,1]}, \qquad (3.5)$$

the mean value of the magnetic dipole moment can be transformed to an algebraic expression depending on the orbital and the spin-isospin quantum numbers of the A-nucleons state and the calculations are, then, straightforward [2].

**4. The Nucleon-Nucleon Interactions**
The nucleon-nucleon interaction $V(|\boldsymbol{r}_i - \boldsymbol{r}_j|)$ of equation (2.1), which has been used in our calculations, has the well-known form [28]

$$V(r) = {}^{ts}X\{V_C(r) + V_T(r)S_{12} + V_{LS}\{r\}\boldsymbol{\ell}.\boldsymbol{s} + V_{LL}(r)L_{12}\}, \qquad (4.1)$$

where $\boldsymbol{\ell}, \boldsymbol{s}$ and $t$ are the orbital angular momentum, the spin angular momentum and the isotopic spin of the two-nucleon state, respectively. The central, tensor, spin-orbit and quadratic spin-orbit terms are standard. The operator ${}^{ts}X$ has the form [28]

$$^{ts}X = C_W + (-1)^{s+t+1}C_M + (-1)^{s+1}C_B + (-1)^{t+1}C_H, \qquad (4.2)$$

where $C_W, C_M, C_B$ and $C_H$ are the Wigner, the Majorana, the Bartlett and the Heisenberg exchange constants, respectively. Each term of the interaction is expressed as a sum of Gaussian functions in the form

$$V_\alpha(r) = \sum_{k=1}^{4} V_{\alpha k}\, e^{-\frac{r^2}{r_{\alpha k}^2}}, \qquad (4.3)$$

where $\alpha = C, T, LS$ and $LL$.

Two sets of values are considered for the exchange constants. For the first set we have: $C_W = 0.1333$, $C_M = -0.9333$, $C_B = -0.4667$ and $C_H = -0.2667$, which are known as the Rosenfeld constants and belong to the symmetric case. We refer to the potential resulting from this case by Pot-I. In the second set we have $C_W = -0.41$, $C_M = -0.41$, $C_B = -0.09$ and $C_H = 0.09$, which belong to the Serber case. The resulting potential is denoted by Pot-II. For the triplet-even state ($t = 0, s = 1$), which is the case for the ground-state of deuteron, and from the normalization condition of the exchange constants the operator ${}^{ts}X$ equals -1, for both of the symmetric and the Serber cases so that the two types of the exchange constants will produce the same results for the ground-state characteristics of deuteron.

In the present paper we have varied the depth and range parameters $V_{\alpha k}$ and $r_{\alpha k}$, respectively, in order to obtain results for the binding energy, the root mean-square radius, the D-state probability, the magnetic dipole moment and the electric quadrupole moment of deuteron in excellent agreement with the corresponding experimental values. The resulting two potentials are then used in our calculations for triton and [4]He.

**5. The Three-Body Forces**
It is well known that three-body forces are important to describe the properties of finite nuclei. The parameters in the nucleon-nucleon potential may not be unique or there may be some redundant parameters in order to reproduce the deuteron properties. In order to investigate these points of view, we have considered the following Hamiltonian operator, which takes into consideration the three-body forces:

$$H = H^{(0)} + V' + V'', \tag{5.1}$$

where the first two terms in (5.1) are given by (2.6), and (2.7) and

$$V'' = \Sigma_{i,j,k} V(\mathbf{r}_i, \mathbf{r}_j, \mathbf{r}_k), \tag{5.2}$$

is the three-body potential. For the three-body potential we have used two potentials: the Skyrme-III potential [29] and the Urbana model (UVII) three nucleon interactions [30]

## 5.1 The Skyrme-III Three-Body Force
The Skyrme-III potential [29] is given by

$$V'' = t_3 \delta(\mathbf{r}_1 - \mathbf{r}_2)\delta(\mathbf{r}_2 - \mathbf{r}_3), \tag{5.3}$$

where $t_3 = 14000.0\ MeV \times fm^6$.

To calculate the matrix elements of the three-body interactions, given by (5.3), we have used the fact that

$$\delta(\mathbf{r}_1 - \mathbf{r}_2) = \frac{1}{r_1 r_2} \delta(r_1 - r_2)\delta(cos\theta_1 - cos\theta_2)\delta(\varphi_1 - \varphi_2). \tag{5.4}$$

For further details see [2].

## 5.2 The Urbana Model Three Nucleon Interactions
Explicit treatments of the pions and deltas degrees of freedom, or implicit treatment via an effective three-body force (3BF), are different ways to approach the same physics. In approach-1, the nucleus is assumed to be consisting of nucleons only. As for the nucleon-nucleon (N-N) forces one uses one of the standard two-body force (2BF) potentials (Reid Soft Core, Paris, Bonn, …), supplemented by a phenomenological or microscopically derived 3BF such as the Tucson-Melbourne (TM) force. This approach is taken by the Los Alamos, Urbana, Tohoku groups [34-36].

The second approach referred to as "3-body force" is based on an explicit treatment of the non-nucleonic degrees of freedom in the ground state wave function. It can be included microscopically by allowing in the wave function for pions, deltas, and pairwise interactions with these additional constituents. This approach is taken by the authors in refs. [37].

The Urbana model (UVII) three nucleon interaction is written as a sum of long-range two pion exchange and intermediate-range repulsive terms [30]

$$V_{ijk} = V_{ijk}^{FM} + V_{ijk}^R = V_{ijk}^{2\pi} + V_{ijk}^R. \tag{5.5}$$

$$V_{ijk}^{FM} = \Sigma_{cyc} -0.0333 \left( \{\tau_i \cdot \tau_j, \tau_i \cdot \tau_k\}\{x_{ij}, x_{ik}\} + \frac{1}{4}[\tau_i \cdot \tau_j, \tau_i \cdot \tau_k][x_{ij}, x_{ik}] \right),$$

$$x_{ij} = T(r_{ij})S_{ij} + \sigma_i \cdot \sigma_j Y(r_{ij}),$$

$$V_{ijk}^R = \Sigma_{cyc} 0.0038\ T^2(r_{ij})T^2(r_{jk}). \tag{5.6}$$

The $T(r)$ and $Y(r)$ are radial functions associated with the tensor and Yukawa parts of the one pion-exchange interaction:

$$Y(r) = \frac{e^{-\mu r}}{\mu r}(1 - e^{-br^2}), \tag{5.7}$$

$$T(r) = \left[1 + \frac{3}{\mu r} + \frac{3}{(\mu r)^2}\right](1 - e^{-br^2}). \tag{5.8}$$

Here the pion mass $\mu = 0.7 fm^{-1}$, and $b = 2 fm^{-2}$. The $V_{ijk}^{2\pi}$ is the familiar Fujita-Miyazawa two-pion exchange operator, and is attractive. The $V_{ijk}^{R}$ is repulsive, and its strength $U_0$ is 0.0038 in model VII instead of 0.003 in model V.

The Coulomb interaction is taken as [38]

$$V_c(r_{ij}) = \frac{e^2}{4r_{ij}}(1 + \tau_{3,i})(1 + \tau_{3,j})\left[1 - \frac{1}{48}e^{-x}(48 + 33x + 9x^2 + x^3)\right], \tag{5.9}$$

and $x = \sqrt{12}r_{ij}/R_{cp}$, where $R_{cp}$ is the rms charge radius of the proton.

## 6. Results and Discussions

In our investigation, the ground- and excited-state wave functions of ³H and ⁴He are expanded in series in terms of the basis functions of the TISM with even number of quanta of excitations $N$ in the range $0 \le N \le 20$. Accordingly, each one of these basis functions is expanded in terms of the two-particle total fractional parentage coefficients, which are products of orbital and spin-isospin coefficients, in order to calculate the two-particle operators of the Hamiltonian. As a result, the Hamiltonian matrices for the different states of ³H and ⁴He are constructed as functions of the oscillator parameter $\hbar\omega$. By diagonalizing these matrices with respect to $\hbar\omega$, as a variational parameter, we obtained the nuclear energy eigenvalues and the corresonding eigenfunctions.

The ground-state of triton has total angular momentum $J = \frac{1}{2}$, isotopic spin $T = \frac{1}{2}$ and even parity, i.e. $(J^\pi, T) = \left(\frac{1^+}{2}, \frac{1}{2}\right)$. In Table-1 we present the used TISM basis functions for the ground-state of triton which correspond to number of quanta of excitations $N = 0, 2, 4, 6, 8, 10, 12, 14, 16, 18$ and $20$. The basis functions which produced in the final calculations weights $\le 10^{-6}$ are eliminated and then the resulting nuclear wave functions are renormalized to unity.

The energy eigenvalues which resulted from the diagonalization of the Hamiltonian matrices for the state $\left(\frac{1^+}{2}, \frac{1}{2}\right)$ of triton for each value of the oscillator parameter $\hbar\omega$, which is allowed to vary in the range $8 \le \hbar\omega \le 20$ MeV, showed two accepted values: the lowest one belongs to the ground-state, and hence the negative value of the binding energy, and the highest belongs to the first-excited state energy $E^*$. Other higher values exist but we did not present them as there is no experimental evidence for these excited states of triton. The obtained ground-state wave functions are used to calculate the root mean-square radius and the magnetic dipole moment of triton.

In Table-2 we present different quantities which characterize the ground state wave function of triton. For this purpose we present in Table-2, the $S$, $D$ and $P$ state probabilities, $P_S$, $P_D$ and $P_P$, respectively, corresponding to bases having total angular momentum $L = 0, 2$ and $1$, respectively, with the different cases of the used potentials. The probabilities of bases having irreducible representations $[f] = [3], [21]$ and $[111]$ of the symmetric group $S_3$ for the ground-state wave functions of triton are also given in this table. Also, the probabilities of bases having $S = \frac{1}{2}$ and $S = \frac{3}{2}$ are given.

In Table-3 we present the triton ground-state wave function $(\Psi)$, after eliminating the bases with weights $\le 10^{-6}$ and then renormalizing $\Psi$, binding energy (B.E.), in MeV, root mean-square radius $(R)$, in fm, first excited–state energy $(E^*)$, in MeV and magnetic dipole moment, $(\mu)$ in N.M., for the two nucleon-nucleon potentials (Pot-I and Pot-II). The improved values which resulted from using the Skyrme III and the UVII three-nucleon interactions are also given in this table. Moreover, the corresponding experimental values and the values of the oscillator parameter $\hbar\omega$ which produced the minimum energy-eigenvalues are also given in Table-3. Furthermore, previous results by using the Faddeev method together with the AV18 nucleon-nucleon interaction and the Tucson-Melbourne (TM) two-pion exchange three-body interaction are also given [39]. Also, previous results [40] by using the NCSM with the AV18 nucleon-nucleon interaction and the Tucson-Melbourne (TM) two-pion exchange three-body interaction are also given in this table.

The even-parity states of the nucleus $^4$He are: the ground-state $(0^+, 0)_1$, and the excited-states $(0^+, 0)_2$, $(2^+, 0)$ and a third state $(1^+, 0)$, which is predicted in [10]. The Hamiltonian matrices which belong to these states are constructed with respect to even number of quanta of excitations $N$ in the range $0 \leq N \leq 20$ as functions of the oscillator parameter $\hbar\omega$, which is allowed to vary in a large range of values $8 \leq \hbar\omega \leq 28$ MeV in order to obtain the best fit to the energies of the even-parity states of $^4$He. In Table-4 we present the TISM basis functions for the ground-state of $^4$He. The basis functions which produced in the final calculations weights $\leq 10^{-6}$ are eliminated and then the resulting nuclear wave functions are renormalized to unity. The eigenvalues which resulted from the diagonalization of the ground state-Hamiltonian matrices for $^4$He showed two accepted values: the lowest one belongs to the ground-state, $(0^+, 0)_1$, and the highest belongs to the first-excited state, $(0^+, 0)_2$. The obtained ground-state nuclear wave functions are used to calculate the root mean-square radius of $^4$He with respect to each value of the oscillator parameter $\hbar\omega$.

In Table-5 we present different quantities which characterize the ground state wave function of $^4$He. Accordingly, we present in Table-5, the $S$, $D$ and $P$ state probabilities: $P_S$, $P_D$ and $P_P$, respectively, for the ground-state wave function of $^4$He by using the two nucleon-nucleon potentials alone. Also, the improved values which resulted from using the two three-nucleon interactions together with the two nucleon-nucleon potentials are given in this table. The probabilities of bases having irreducible representations $[f] = [4], [31], [22]$ and $[211]$ of the symmetric group $S_4$ for the ground-state wave functions of $^4$He are also given.

In Table-6 we present the ground-state wave function of $^4$He ($\Psi$), after eliminating the bases with weights $\leq 10^{-6}$ and then renormalizing $\Psi$, the binding energy (B.E.), in MeV, the even-parity excited-state energies, in MeV and the root mean-square radius ($R$), in fm, for the two nucleon-nucleon potentials (Pot-I and Pot-II) in the case of $N = 0, 2, 4, 6, 8, 10, 12, 14, 16, 18$ and $20$. The improved values which resulted from using the Skyrme III and the UVII three-nucleon interactions are also given in this table. Moreover, the corresponding experimental values and the values of the oscillator parameter $\hbar\omega$ which produced the best fit between the calculated energies of the even parity-states of $^4$He and the correspondoing experimental values are also given in Table-6. Furthermore, previous results by using the Faddeev and Yakubovsky method [25] and the NCSM method [25] are also given in this table.

It is seen from Tables-3 and -6 that the inclusion of the three-body interactions improved the calculated ground-state as well as the excited state characteristics of triton and $^4$He for the two potentials, as expected. Also, it is seen from Tables-3 and -6 that the calculated values of the different characteristics of triton and $^4$He by using Pot-I together with UVII three-body interaction are in excellent agreement with the corresponding experimental values rather than the other cases.

The variations of the triton binding energy (B.E.), root mean-square radius ($R$), first excited–state energy ($E^*$) and magnetic dipole moment of triton ($\mu$) with respect to the oscillator parameter $\hbar\omega$ (i.e. the dependence of the obtained results on the used model and its wave function) for the two potentials, with and without improvements arised from using the two three-body interactions, are given in Figs. 1-8. Moreover, the variations of the $^4$He binding energy (B.E.), root mean-square radius ($R$) and first excited–state energy ($E^*$) with respect to the oscillator parameter $\hbar\omega$ for the two potentials, with and without improvements arised from using the two three-body interactions, are given in Figs. 9-14. The figures show minima for the values of the root mean-square radius, the ground-stae energy (the negative value of the binding energy) and the first-excited state energy for the helium nucleus, for the two cases of the potentials, in agreement with the basic property of the used model. For the triton nucleus minima have been obtained only for the ground-state and the root-mean square radius.

Concerning the second $\left(\frac{1^+}{2}, \frac{1}{2}\right)$ state of $^3$H, we would like to point out that the three- and four-nucleon states have only one bound state, and all excited states are in the continuum. The use of a bound-state approach with square-integrable basis function is only meaningful for bound states, and narrow resonances. Contrary to $^4$He, which presents a broad $0^+$ resonance, there are no resonances in the $\frac{1^+}{2}$ channel of tritium. The reason for which its energy increases with increasing harmonic oscillator frequency and does not present a clear minimum is that this eigenstate and those above it represent a discretization of the energy continuum and as such the continuously move as the model space size is increased or other parameters are varied.

It is well-known that calculations within a three-body translational invariant harmonic oscillator basis and using realistic two and three-nucleon forces have been performed for the three-nucleon system (see e.g. [40]), and for the four-nucleon system (see e.g. [43,44]). These were obtained within the translational invariant form of the no-core shell model, which is completely equivalent to the TISM with the exception that the antisymmetrization of the wave function is not achieved by means of group theory, but rather by diagonalizing the antisymmetrization operator and retaining as basis states the antisymmetric eigenstates. It is well-known also that, the three- and four-nucleon systems have been studied by means of numerically exact few-body approaches (such as the Faddeev, Faddeev-Yacubosky and hyperspherical harmonics approaches starting from realistic two and three-nucleon forces, see for example [25, 39, 45]). A review can be found e.g., in [46].

However, the calculations presented here does present a certain degree of novelty with respect to the translational invariant no-core shell model. In particular, the direct construction of antisymmetric three- and four-body basis states with the help of group theory is more elegant and may even turn out to be computationally more advantageous. Moreover, in order to study the convergence properties of the TISM method with respect to the dimension of the adopted model space, the calculations with the TISM have been extrapolated by using the Stirling's formula [47] for the results with $N \leq 20$, step-by- step, to reach $N = 30$ for the two nuclei. In Figures 15-18 we present the dependence of the binding energy (B.E.), root mean-square radius ($R$), first excited-state energy ($E^*$) and magnetic dipole moment ($\mu$) of triton on the number of quanta of excitations $N$, respectively, by using Pot-I + Skyrme, Pot-I + UVII, Pot-II + Skyrme and Pot-II + UVII. Furthermore, we present in figures 19, 20 and 21 the dependence of the binding energy, root mean-square radius and first excited-state energy of $^4$He on the number of quanta of excitations $N$, respectively, by using Pot-I + Skyrme, Pot-I + UVII, Pot-II + Skyrme and Pot-II + UVII.

Finally, we have seen that the group theoretical methods which have been used in this paper enabled us to construct total antisymmetric nuclear wave functions for the ground as well as the excited states of the $^3$H and $^4$He nuclei which take into consideration all the possible configurations and symmetries of the nucleus as a whole and of its individual nucleons. Also, these methods enabled us to calculate the various one-, two- and three-body matrix elements of orbital, spin and isospin operators by means of the one- and two-particle orbital and spin-isospin fractional parentage coefficients. Accordingly, these wave functions can be used in the calculations of the ground- as well as the excited-stae characteristics of nuclei. Other nuclear characteristics such as the mean lifetime of the β-decay [8] and the integral cross section of the $\gamma$-quanta photo absorption by nuclei [10] can be also calculated with such wave functions. Moreover, the convergence of the calculated characteristics of the ground and the excited states of $^3$H and $^4$He to their experimental values is very clear especially for the binding energy and the root mean-square radius of the two nuclei.

Table-1 TISM-basis functions for the triton nucleus with $0 \leq N \leq 20$. The basis functions which produced in the final calculations weights $\leq 10^{-6}$ are eliminated.

| $\Psi_i$ i | N | {ρ} | (ν) | [f] | L | S |
|---|---|---|---|---|---|---|
| 1 | 0 | {0} | (0) | [3] | 0 | 1/2 |
| 2 | 2 | {2} | (0) | [3] | 0 | 1/2 |
| 3 | 2 | {2} | (2) | [21] | 0 | 1/2 |
| 4 | 2 | {2} | (2) | [21] | 2 | 3/2 |
| 5 | 2 | {11} | (0)* | [$1^3$] | 1 | 1/2 |
| 6 | 4 | {4} | (0) | [3] | 0 | 1/2 |
| 7 | 4 | {4} | (2) | [21] | 0 | 1/2 |
| 8 | 4 | {4} | (2) | [21] | 2 | 3/2 |
| 9 | 4 | {31} | (2) | [21] | 1 | 1/2 |
| 10 | 6 | {6} | (0) | [3] | 0 | 1/2 |
| 11 | 6 | {6} | (2) | [21] | 0 | 1/2 |
| 12 | 8 | {8} | (0) | [3] | 0 | 1/2 |
| 13 | 10 | {10} | (0) | [3] | 0 | 1/2 |
| 14 | 12 | {12} | (0) | [3] | 0 | 1/2 |
| 15 | 14 | {14} | (0) | [3] | 0 | 1/2 |
| 16 | 16 | {16} | (0) | [3] | 0 | 1/2 |
| 17 | 18 | {18} | (0) | [3] | 0 | 1/2 |
| 18 | 20 | {20} | (0) | [3] | 0 | 1/2 |

Table-2 The $S$, $D$ and $P$ state probabilities, $P_S$, $P_D$ and $P_P$, respectively, for the ground-state wave functions of triton. The probabilities of bases having irreducible representations $[f] = [3], [21]$ and $[111]$ of the symmetric group $S_3$ for the ground-state wave functions of triton are also given. The probabilities of bases having $S = \frac{1}{2}$ and $S = \frac{3}{2}$ are also given.

| Case Characteristic | Pot-I | Pot-I + Skyrme | Pot-I + UVII | Pot-II | Pot-II + Skyrme | Pot-II + UVII |
|---|---|---|---|---|---|---|
| $P_S\%$ | 93.869 | 92.806 | 92.065 | 92.557 | 94.195 | 93.961 |
| $P_D\%$ | 5.910 | 6.974 | 7.348 | 6.999 | 5.344 | 5.927 |
| $P_P\%$ | 0.221 | 0.220 | 0.587 | 0.444 | 0.461 | 0.112 |
| $P_{[3]}\%$ | 85.782 | 85.941 | 85.901 | 85.815 | 89.544 | 90.126 |
| $P_{[21]}\%$ | 14.178 | 14.020 | 13.888 | 14.118 | 10.386 | 9.826 |
| $P_{[111]}\%$ | 0.040 | 0.038 | 0.210 | 0.067 | 0.069 | 0.048 |
| $P_{S=\frac{1}{2}}\%$ | 94.090 | 93.026 | 92.652 | 93.000 | 94.655 | 94.073 |
| $P_{S=\frac{3}{2}}\%$ | 5.910 | 6.974 | 7.348 | 6.999 | 5.344 | 5.927 |

Table-3 The $^3$H ground-state wave function, the binding energy (B.E.), in MeV, the first excited-stae enegy ($E^*$), in MeV, the root mean-square radius ($R$), in fm, and the magnetic dipole moment ($\mu$), in N.M., by using the two nucleon-nucleon potentials (Pot-I and Pot-II) and also the improved results by using the Skirme III and the UVII three-nucleon interactions. The values of $\hbar\omega$, which gave the best fit to the binding energy of $^3$H are also given. Previous results by using the Faddeev method together with the AV18 nucleon-nucleon interaction and the Tucson-Melbourne (TM) two-pion exchange three-body interaction are also given [39]. Also, previous results [40] by using the NCSM with the same interactions in ref. [39] are also given.

|  | Pot-I | Pot-I + Skyrme | Pot-I + UVII | Pot-II | Pot-II + Skyrme | Pot-II + UVII | Faddeev AV18+ TM [39] | NCSM, AV18+ TM [40] | Exper. |
|---|---|---|---|---|---|---|---|---|---|
| $\psi_1$ | 0.8925 | 0.8972 | 0.9008 | 0.9002 | 0.8966 | 0.9034 | --- | --- | --- |
| $\psi_2$ | 0.1113 | 0.0796 | 0.1549 | 0.0960 | 0.2242 | 0.1917 | --- | --- | --- |
| $\psi_3$ | -.2434 | -.2223 | -.2087 | -.2222 | 0.1718 | -.1034 | --- | --- | --- |
| $\psi_4$ | 0.2125 | 0.2338 | 0.2273 | 0.2229 | 0.1770 | 0.2061 | --- | --- | --- |
| $\psi_5$ | 0.0200 | 0.0195 | 0.0458 | 0.0258 | 0.0262 | 0.0218 | --- | --- | --- |
| $\psi_6$ | 0.1964 | 0.1952 | 0.1244 | 0.1750 | 0.1735 | 0.1953 | --- | --- | --- |
| $\psi_7$ | 0.1437 | 0.1353 | 0.1321 | 0.1321 | 0.1283 | 0.1397 | --- | --- | --- |
| $\psi_8$ | 0.1181 | 0.1228 | 0.1477 | 0.1425 | 0.1487 | 0.1296 | --- | --- | --- |
| $\psi_9$ | 0.0425 | 0.0427 | 0.0614 | 0.0613 | 0.0626 | 0.0253 | --- | --- | --- |
| $\psi_{10}$ | 0.0130 | 0.0093 | 0.0310 | 0.0210 | 0.0161 | 0.0087 | --- | --- | --- |
| $\psi_{11}$ | -.0312 | -.0303 | -.0247 | -.0247 | -.0228 | -.0902 | --- | --- | --- |
| $\psi_{12}$ | 0.0519 | 0.0516 | 0.0447 | 0.0447 | 0.0563 | 0.0514 | --- | --- | --- |
| $\psi_{13}$ | 0.0491 | 0.0421 | 0.0313 | 0.0313 | 0.0410 | 0.0422 | --- | --- | --- |
| $\psi_{14}$ | 0.0377 | 0.0411 | 0.0299 | 0.0299 | 0.0334 | 0.0399 | --- | --- | --- |
| $\psi_{15}$ | 0.0352 | 0.0363 | 0.0310 | 0.0310 | 0.0303 | 0.0370 | --- | --- | --- |
| $\psi_{16}$ | 0.0338 | 0.0341 | 0.0364 | 0.0364 | 0.0486 | 0.0342 | --- | --- | --- |
| $\psi_{17}$ | 0.0289 | 0.0293 | 0.0228 | 0.0298 | 0.0325 | 0.0313 | --- | --- | --- |
| $\psi_{18}$ | 0.0197 | 0.0213 | 0.0214 | 0.0214 | 0.0246 | 0.0252 | --- | --- | --- |
| B.E. | 8.2915 | 8.4794 | 8.4797 | 8.2173 | 8.4311 | 8.4457 | 8.444 | 8.39 | 8.48 [41] |
| $E^*$ | 8.7431 | 8.5212 | 8.5015 | 8.7754 | 8.6654 | 8.6341 | --- | 6.2065 | --- |
| $R$ | 1.8062 | 1.7512 | 1.7509 | 1.9240 | 1.7729 | 1.7644 | --- | --- | 1.75 [41] |
| $\mu$ | 3.303 | 3.281 | 3.259 | 3.562 | 3.392 | 3.369 | --- | --- | 2.98 [2] |
| $\hbar\omega$ | 14 | 14 | 14 | 14 | 14 | 14 | --- | --- | --- |

Table-4 The TISM bases of the ground-state of $^4$He. The basis functions which produced in the final calculations weights $\leq 10^{-6}$ are eliminated.

| $\Psi_i$ $i$ | $N$ | $\{\rho\}$ | $(\nu)$ | $[f]$ | $L$ | $S$ |
|---|---|---|---|---|---|---|
| 1 | 0 | {0} | (0) | [4] | 0 | 0 |
| 2 | 2 | {2} | (0) | [4] | 0 | 0 |
| 3 | 2 | {2} | (0) | [22] | 0 | 0 |
| 4 | 2 | {2} | (2) | [22] | 2 | 2 |
| 5 | 2 | {11} | (1) | [211] | 1 | 1 |
| 6 | 4 | {4} | (0) | [4] | 0 | 0 |
| 7 | 4 | {4} | (2) | [22] | 0 | 0 |
| 8 | 4 | {4} | (2) | [22] | 2 | 2 |
| 9 | 4 | {31} | (2) | [31] | 1 | 1 |
| 10 | 6 | {6} | (0) | [4] | 0 | 0 |
| 11 | 6 | {6} | (2) | [22] | 2 | 2 |
| 12 | 8 | {8} | (0) | [4] | 0 | 0 |
| 13 | 8 | {8} | (2) | [22] | 2 | 2 |
| 14 | 10 | {10} | (0) | [4] | 0 | 0 |
| 15 | 10 | {10} | (2) | [22] | 2 | 2 |
| 16 | 12 | {12} | (0) | [4] | 0 | 0 |
| 17 | 14 | {14} | (0) | [4] | 0 | 0 |
| 18 | 16 | {16} | (0) | [4] | 0 | 0 |
| 19 | 18 | {18} | (0) | [4] | 0 | 0 |
| 20 | 20 | {20} | (0) | [4] | 0 | 0 |

Table-5 The $S$, $D$ and $P$ state probabilities, $P_S$, $P_D$ and $P_P$, respectively, for the ground-state wave function of $^4$He for the different cases of the used potentials. The probabilities of bases having irreducible representations $[f] = [4], [31], [22]$ and $[211]$ of the symmetric group $S_4$ are also given.

| Case Characteristic | Pot-I | Pot-I + Skyrme | Pot-I + UVII | Pot-II | Pot-II + Skyrme | Pot-II + UVII |
|---|---|---|---|---|---|---|
| $P_S$% | 90.02 | 92.98 | 86.7 | 91.230 | 89.6 | 87.25 |
| $P_D$% | 7.82 | 5.23 | 11.90 | 6.44 | 6.75 | 9.08 |
| $P_P$% | 2.16 | 1.79 | 1.4 | 2.33 | 3.65 | 3.67 |
| $P_{[4]}$% | 89.53 | 91.548 | 85.695 | 91.198 | 87.606 | 87.22 |
| $P_{[31]}$% | 1.962 | 1.517 | 0.2937 | 1.8009 | 2.3716 | 3.663 |
| $P_{[22]}$% | 8.288 | 6.657 | 12.95 | 6.475 | 8.699 | 9.1026 |
| $P_{[211]}$% | 0.197 | 0.2714 | 1.056 | 0.524 | 1.2746 | 0.011 |

Table-6 The ground-state wave function of $^4$He ($\Psi$), after eliminating the bases with weights $\leq 10^{-6}$ and then renormalizing $\Psi$, the binding energy (B.E.), in MeV, the even-parity excited-state energies, in MeV and the root mean-square radius ($R$), in fm, for the two potentials (Pot-I and Pot-II). The improved values which resulted from using the Skyrme III and the UVII three-nucleon interactions are given. The corresponding experimental values and the values of the oscillator parameter $\hbar\omega$, in MeV, which gave the best fit between the calculated energies of the even parity-states of $^4$He and the correspondoing experimental values are also given. Previous results by using the Faddeev and Yakubovsky (FY) method [25] and the NCSM method [25] are also given in this table.

|  | Pot-I | Pot-I + Skyrme | Pot-I + UVII | Pot-II | Pot-II + Skyrme | Pot-II + UVII | FY [25] | NCSM [25] | Exper. |
|---|---|---|---|---|---|---|---|---|---|
| $\psi_1$ | 0.9282 | 0.9372 | 0.9088 | 0.9254 | 0.8764 | -0.9121 | --- | --- | --- |
| $\psi_2$ | -0.0876 | -0.0916 | -0.0911 | -0.0768 | 0.1528 | -0.1255 | --- | --- | --- |
| $\psi_3$ | 0.0421 | 0.0755 | 0.0624 | 0.0145 | -0.0671 | 0.0136 | --- | --- | --- |
| $\psi_4$ | -0.2213 | -0.1821 | -0.2746 | -.2082 | 0.1723 | -0.2134 | --- | --- | --- |
| $\psi_5$ | 0.0444 | 0.0521 | 0.1028 | 0.0724 | -0.1129 | 0.0105 | --- | --- | --- |
| $\psi_6$ | -0.1174 | 0.1204 | -0.1178 | -.0742 | 0.1784 | -0.0744 | --- | --- | --- |
| $\psi_7$ | 0.0544 | 0.0926 | 0.0812 | 0.0097 | 0.1223 | 0.0077 | --- | --- | --- |
| $\psi_8$ | -0.0982 | -.0548 | -0.1174 | -.0778 | -0.0920 | -0.1344 | --- | --- | --- |
| $\psi_9$ | 0.1401 | 0.1232 | 0.0542 | 0.1342 | 0.1540 | 0.1914 | --- | --- | --- |
| $\psi_{10}$ | 0.0104 | 0.0019 | 0.0097 | 0.0842 | 0.0930 | -0.0634 | --- | --- | --- |
| $\psi_{11}$ | 0.0911 | 0.0810 | 0.1106 | 0.0956 | -0.0997 | 0.0848 | --- | --- | --- |
| $\psi_{12}$ | 0.0098 | 0.0211 | 0.0103 | 0.1156 | 0.1305 | -0.0892 | --- | --- | --- |
| $\psi_{13}$ | -0.0843 | -0.0932 | -0.0982 | 0.0582 | 0.0971 | 0.0768 | --- | --- | --- |
| $\psi_{14}$ | 0.0714 | 0.0665 | 0.0677 | 0.1111 | -0.1453 | 0.0067 | --- | --- | --- |
| $\psi_{15}$ | -0.0643 | -0.0298 | -0.0892 | -.0502 | -0.1001 | -0.1187 | --- | --- | --- |
| $\psi_{16}$ | 0.0111 | 0.0107 | 0.0161 | 0.0109 | 0.0453 | 0.0786 | --- | --- | --- |
| $\psi_{17}$ | 0.0225 | 0.0167 | 0.0183 | 0.0177 | 0.0184 | 0.0159 | --- | --- | --- |
| $\psi_{18}$ | 0.0421 | 0.0340 | 0.0386 | 0.0516 | 0.0443 | 0.0164 | --- | --- | --- |
| $\psi_{19}$ | 0.0675 | 0.0884 | 0.0444 | 0.0911 | 0.0346 | 0.0122 | --- | --- | --- |
| $\psi_{20}$ | 0.0077 | 0.0044 | 0.0052 | 0.0052 | 0.0218 | 0.0118 | --- | --- | --- |
| $\hbar\omega$ | 17 | 17 | 17 | 17 | 17 | 17 |  |  | --- |
| B.E. | 27.895 | 28.286 | 28.289 | 27.554 | 28.074 | 28.078 | 25.94 | 25.80 | 28.3 [42] |
| $(0^+,0)_2$ | 20.924 | 20.354 | 20.267 | 21.515 | 21.422 | 20.984 | ---- | --- | 20.21 [42] |
| $(2^+,0)$ | 29.494 | 29.212 | 28.413 | 29.556 | 29.373 | 28.876 | ---- | ---- | 28.0 [42] |
| $(1^+,0)$ | 31.615 | 30.821 | 30.142 | 31.662 | 30.876 | 30.337 | ---- | ---- | --- |
| $R$ | 1.534 | 1.468 | 1.464 | 1.687 | 1.512 | 1.493 | 1.485 | 1.485 | 1.46 [42] |

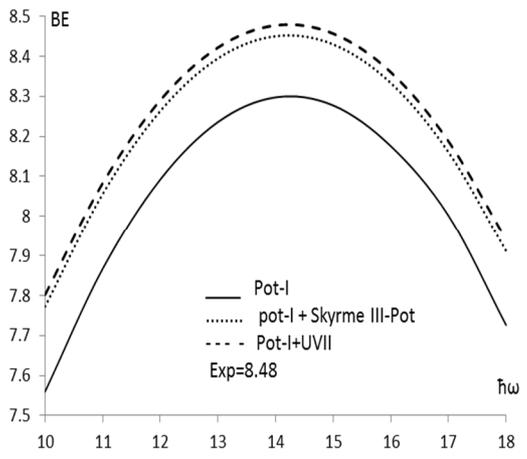
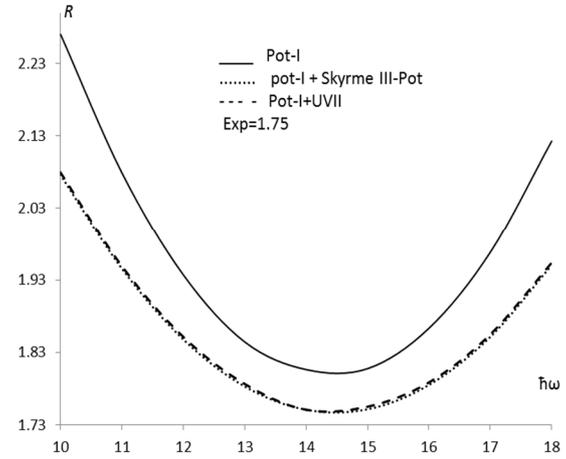

Fig. 1                    Fig. 2

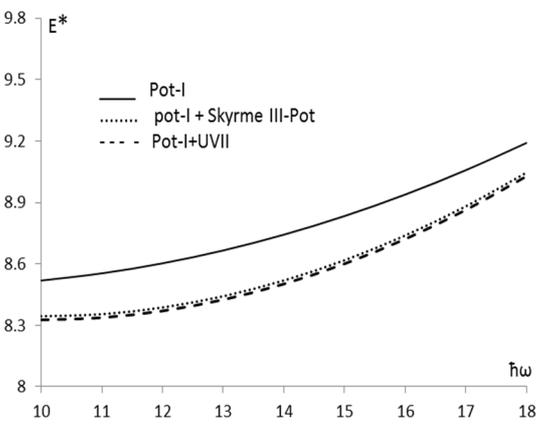
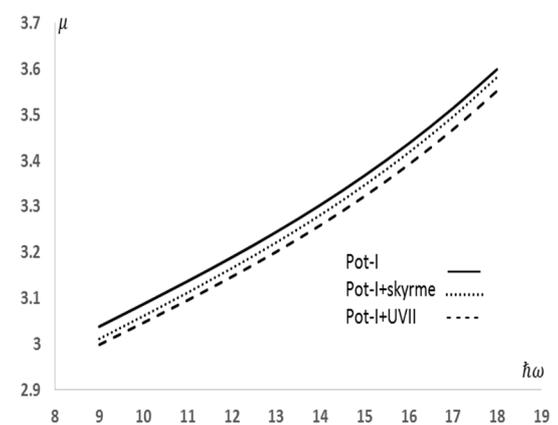

Fig. 3                    Fig. 4

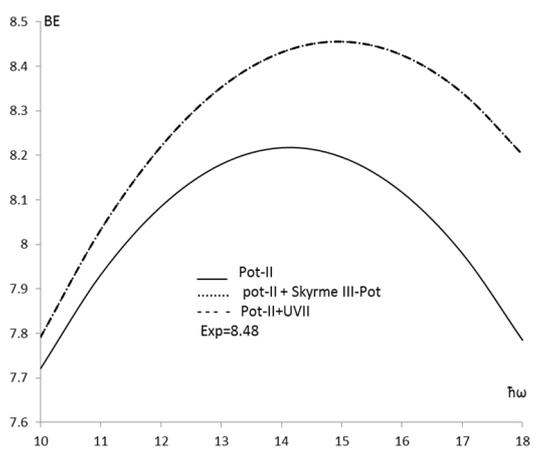
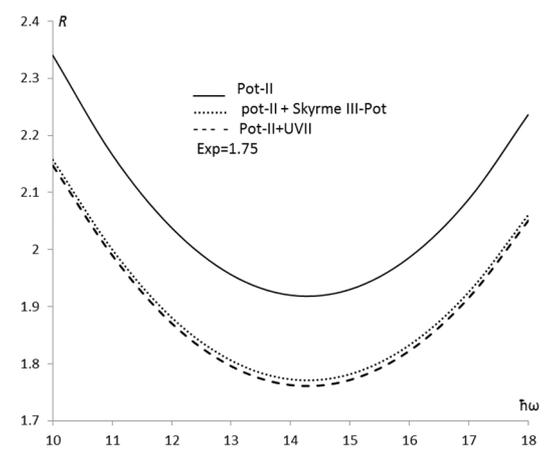

Fig. 5                    Fig. 6

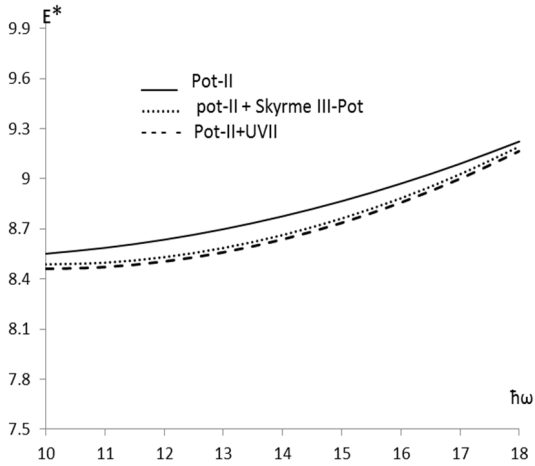

Fig. 7

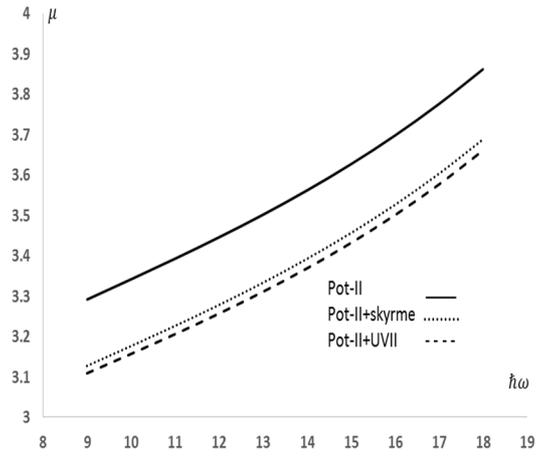

Fig. 8

Figs.1, 2, 3, 4, 5, 6, 7 and 8 Variations of the $^3$H binding energy (B.E.), root mean-square radius ($R$), first excited–state energy ($E^*$) and magnetic dipole moment ($\mu$) with respect to the oscillator parameter $\hbar\omega$ for each potential (Pot-I and Pot-II), respectively. The improved results arising from using the Skyrme III and the UVII three-body interactions are also given.

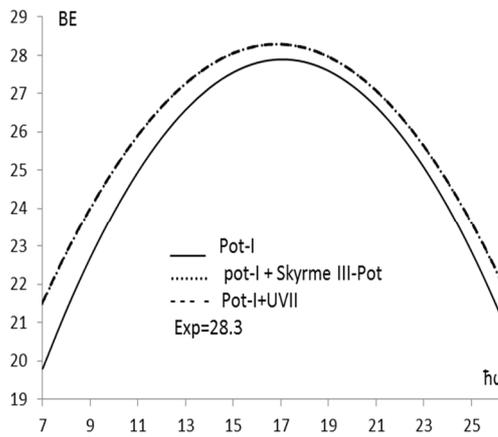

Fig. 9

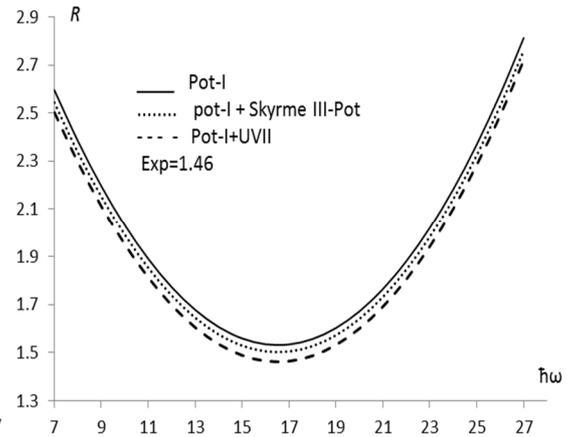

Fig. 10

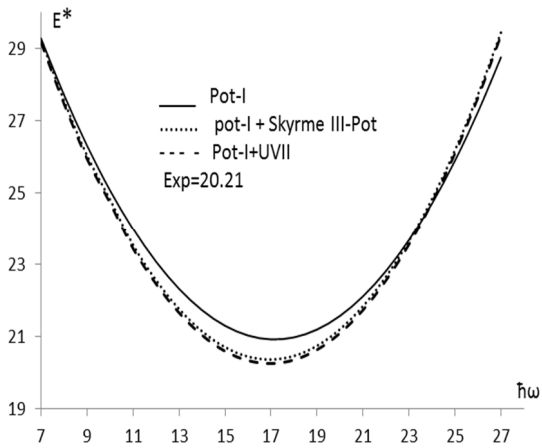

Fig. 11

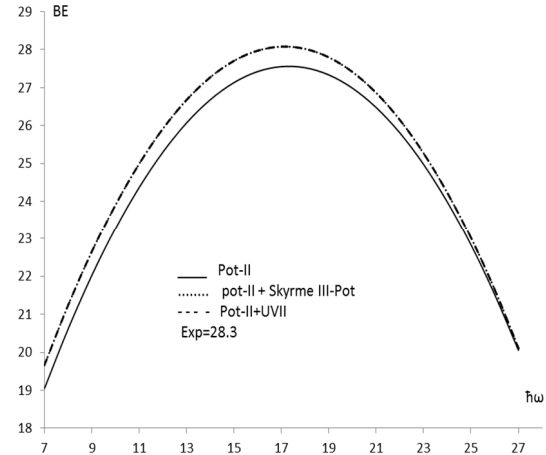

Fig. 12

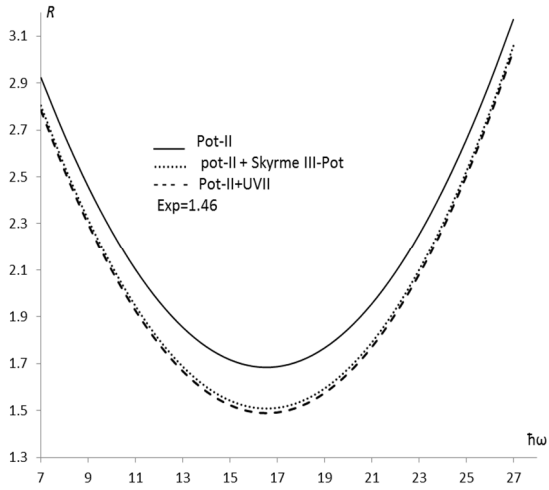 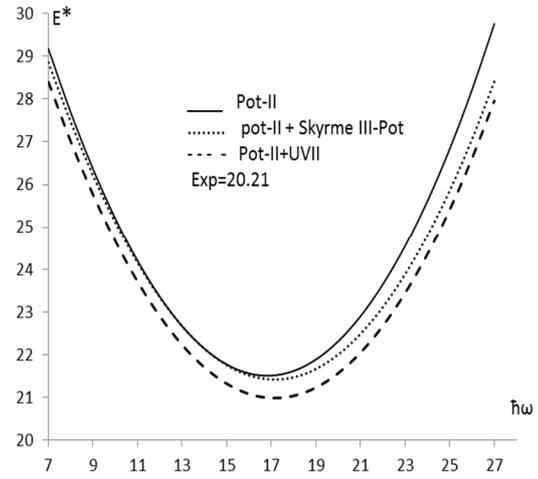

Fig. 13  Fig. 14

Figs. 9, 10, 11, 12, 13, and 14. Variations of the binding energy (B.E.), root mean-square radius ($R$), and first excited–state energy ($E^*$), of $^4$He with respect to the oscillator parameter $\hbar\omega$ for each potential (Pot-I and Pot-II), respectively. The improved results arising from using the Skyrme III and the UVII three-body interactions are also given.

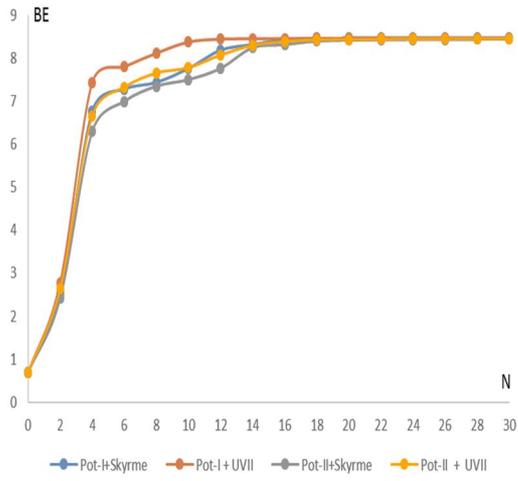

Fig. 15

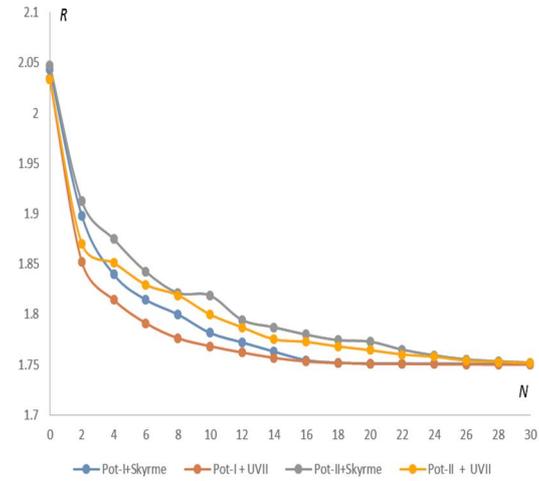

Fig. 16

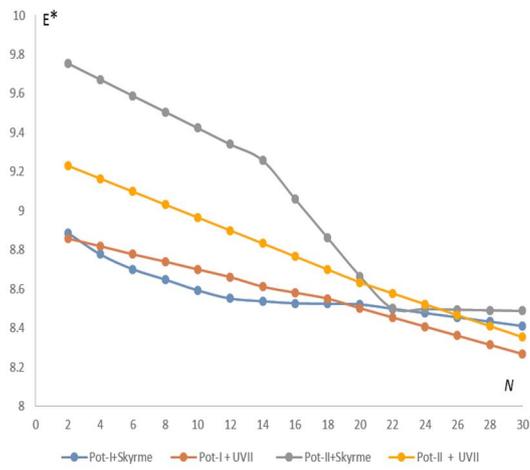

Fig. 17

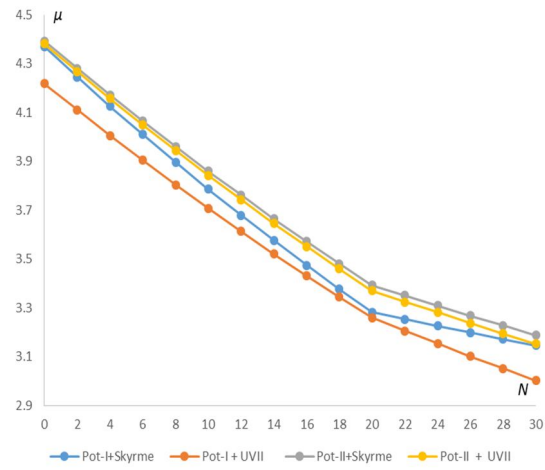

Fig. 18

Figs. 15, 16, 17 and 18. Dependence of the binding energy (B.E.), root mean-square radius ($R$), first excited-state energy ($E^*$) and magnetic dipole moment ($\mu$) of triton on the number of quanta of excitations $N$, respectively, by using Pot-I + Skyrme, Pot-I + UVII, Pot-II + Skyrme and Pot-II + UVII.

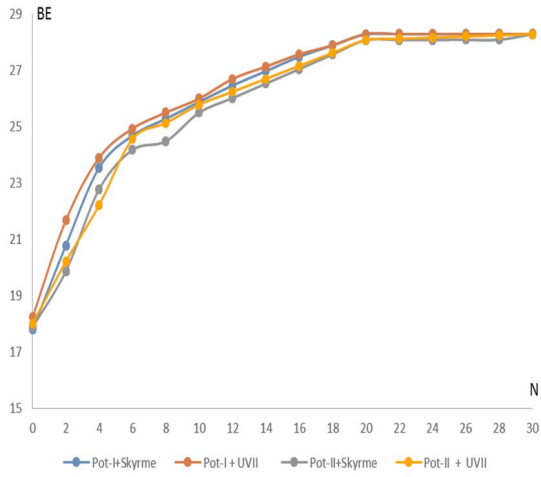
Fig. 19

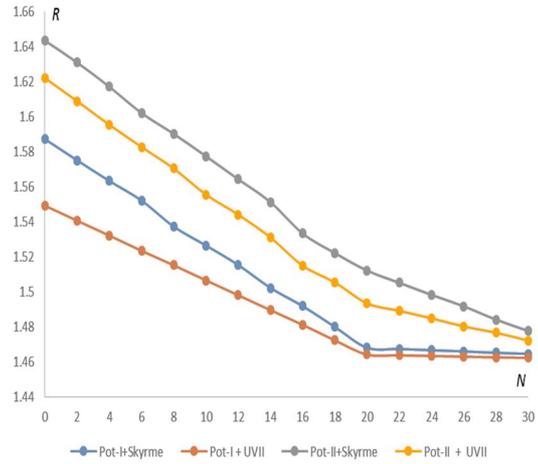
Fig. 20

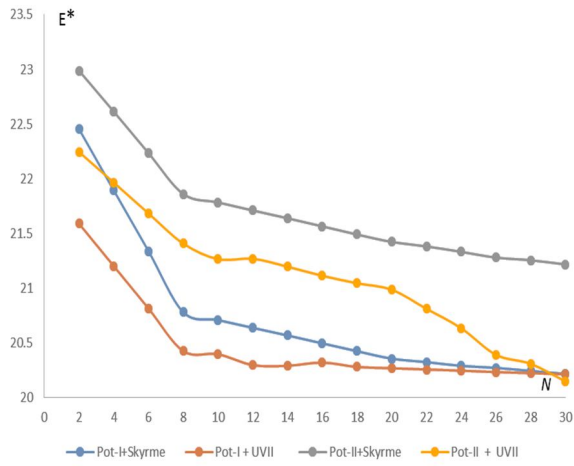
Fig.21

Figs. 19, 20 and 21. Dependence of the binding energy, root mean-square radius and first excited-state energy of $^4$He on the number of quanta of excitations $N$, respectively, by using Pot-I + Skyrme, Pot-I + UVII, Pot-II + Skyrme and Pot-II + UVII.